\documentclass{svjour2}                    
\smartqed  
\usepackage{graphicx,amssymb}
\begin{document}

\title{Quark matter imprint on Gravitational Waves from oscillating
stars}
\titlerunning{Quark matter imprint}
\authorrunning{O. Benhar et al.}
\author{Omar Benhar \and Valeria Ferrari \and Leonardo Gualtieri \and
        Stefania Marassi }

\institute{ Dipartimento di Fisica ``G. Marconi'', Universit\'a degli
Studi di Roma, ``La Sapienza'' \and INFN, Sezione Roma 1, P.le A. Moro
2, 00185 Roma, Italy}

\date{}
\maketitle

\begin{abstract}
We discuss the 
possibility that the detection of gravitational waves emitted by 
compact stars may allow to  constrain  the MIT bag model
of quark matter equation of state.  Our results show that the combined
knowledge of the frequency of the emitted gravitational wave and of
the mass, or the radiation radius, of the source allows one to
discriminate between strange stars and neutron stars and set
stringent bounds on the bag constants.
\end{abstract}


What is the absolute ground state of matter? A definite answer to
this fundamental question is still lacking. Following Bodmer's
seminal paper \cite{Bodmer}, in 1980s Witten \cite{Witten}
suggested that matter consisting of degenerate up, down and strange
quarks, called ``strange'' quark matter,  might be bound and stable
at zero temperature and pressure. This hypotesis,  while not being
directly verifiable in terrestrial laboratories, may be confirmed by
the observation of ``strange stars'', i.e. compact astrophysical
objects entirely made of strange matter except, possibly, for a thin
outer crust.

Theoretical studies of structural properties of quark stars, pioneered
by Itho over three decades ago \cite{Itoh70}, have been booming after
ROSAT discovered the isolated pulsar RXJ1856.5-3754 \cite{ROSAT}. The
data reported in Ref. \cite{Drake} seemed in fact to indicate that its
radiation radius should be in the range $3.8-8.2$~km, so that the
corresponding stellar radius would be far too small compared to that
typical of neutron stars.  The results of this analysis triggered a
number of speculations on the nature of RXJ1856.5-3754, and it 
was suggested that it may be a strange star, though this hypotesis
has now been ruled out \cite{TurollaTrumperPons}.

Whether quark stars do exist in nature is an open question, and the
astrophysical scenario in which they may form is still poorly
understood. For instance, it has been argued that, besides the
standard gravitational collapse, strange stars may form in low-mass
X-ray binaries when, due to accretion, matter in a neutron star core
reaches sufficiently high densities to undergo a deconfinement phase
transition to quark matter \cite{Cheng}.

In this letter we discuss the possibility that detection of a
gravitational signal emitted by a compact star, oscillating in its
fundamental mode with frequency $\nu_f$ and damping time $\tau_f$,
will allow to infer whether the source is a neutron star or a
strange star and to constrain theoretical models of the quark matter
equation of state (EOS).

Due to the complexity of the fundamental theory of strong interactions
between quarks (Quantum Chromo-Dynamics, or QCD), theoretical studies
of strange stars are necessarily based on models. The most used is the
MIT bag model \cite{bagmodel}, in which the two main elements of
QCD, namely color confinement and asymptotic freedom, are implemented
through the assumptions that: i) quarks occur in color neutral
clusters confined to a finite region of space (the bag), the volume of
which is limited by the pressure of the QCD vacuum (the bag constant
$B$), and ii) residual interactions between quarks are weak, and can
be treated in low order perturbation theory in the color coupling
constant $\alpha_s$.

The bag model parameters are the bag constant $B$, the quark masses
$m_f$ (the index $f = u,d,s$ labels the three active quark flavors; at
the densities relevant to our work heavier quarks do not play a role)
and the running coupling constant $\alpha_s$, whose value at the
energy scale $\mu$ can be obtained from the renormalization group
relation
\begin{eqnarray}
\alpha_s(\mu)=\frac{12\pi}{(33-2N_f)\ln(\mu^2/\Lambda^2_{QCD})} \ .
\label{lambda}
\end{eqnarray}
In the above equation $N_f=3$ denotes the number of active flavors,
while $\Lambda_{QCD}$ is the QCD scale parameter, whose value is
constrained to the range $100-250$~MeV by high energy data (see, e.g.,
\cite{QCD_book}).  At the scale typical of the quark chemical
potentials Eq. (\ref{lambda}) yields
\begin{equation}
\label{alfas}
\alpha_s \in [0.4,0.6] \ .
\end{equation}

As quarks are not observable as individual particle, their masses are
not directly measurable. However, they can be inferred from hadron
properties, supplemented by theoretical calculations. According to the
2002 Edition of the Review of Particle Physics \cite{PDG}, the masses
of up and down quarks do not exceed few MeV, and can therefore be
safely neglected, while the mass of the strange quark is much larger,
its value being in the range
\begin{equation}
\label{ms}
                      m_s \in  [80,155] ~ {\rm MeV}\ .
\end{equation}
The bag constant is subject to a much larger uncertainty.  In early
applications of the MIT bag model $B$, $\alpha_s$ and $m_s$ were
adjusted to fit the measured properties of light hadrons (spectra,
magnetic moments and charge radii). This procedure leads to values of
$B$ that differ from one another by as much as a factor of $\sim$ 6,
ranging from $57.5$ MeV/fm$^3$ \cite{DeGrand75} to $351.7$ MeV/fm$^3$
\cite{Carlson83}, while the corresponding values of $\alpha_s$ turn
out to be close to or even larger than unity.  Moreover, the strange
quark masses resulting from these analyses are typically much larger
than the upper limit given by Eq. (\ref{ms}).  Using the parameters
determined from fits to light hadron spectra to describe bulk quark
matter is questionable, as these spectra are known to be strongly
affected by the details of the bag wave-functions, as well as by
spurious contributions arising from the center of mass motion.

The requirement that strange quark matter be absolutely stable at zero
temperature and pressure implies that $B$ cannot exceed the maximum
value $B_{max}\,\approx 95$\,MeV/fm$^3$~\cite{Farhi84}.  For
values of $B$ exceeding $B_{max}$, a star entirely made of deconfined
quarks is not stable. Under these conditions quark matter can only
occupy a fraction of the available volume and the star is said to be
hybrid \cite{glend2}. The results of Ref. \cite{Akmal,Rubino}, based on a 
state-of-the art description of the low-density hadronic phase, 
suggest that, assuming that the transition to quark matter proceeds
through the formation of a mixed phase, hybrid stars can only exist
for values of the bag constant up to $\sim 200$ MeV/fm$^3$ and 
contain a rather small amount of quark matter. Gravitational emission
from the hybrid star models of Ref. \cite{Akmal}, corresponding to 
$m_s = 150$ MeV, $\alpha_s = 0.5$ and $B=120$ and $200$\,MeV/fm$^3$, referred to as 
APRB120 and APRB200, respectively, has been analyzed  in Ref. \cite{astero}.

In this paper we focus on bare strange stars and consider values of
the bag constant in the range
\begin{equation}
\label{B1}
B \in [57,95]~ {\rm MeV}/{\rm fm}^3 \ .
\end{equation}

The problem we shall investigate is the following.
Suppose that a gravitational signal is detected, which is emitted by a
compact object oscillating in its fundamental mode ($\bf f$-mode), 
i.e.  the mode which is known to be the most efficient as far as gravitational
radiation is concerned \cite{Allen98}. We do not know whether the
source is a neutron star or a strange star.
Since the damping time of the $\bf f$-mode , $\tau_f$, is known to be of the
order of a fraction of second, the mode excitation would correspond to
a sharp peak in the energy spectrum of the detected signal, emitted 
at the mode frequency $\nu_f$ which could be identifiable by a suitable 
data analysis technique.
The questions we want to address are:

\begin{itemize}
\item
Does the knowledge of  $\nu_f$ (and/or of $\tau_f$) allow one to
say anything about the nature of the source?
\item
Assuming that we can establish the star  is a strange star, would
these data allow one to set constraints on the parameters of the MIT
bag model?
\end{itemize}
To answer these questions, we compute frequency and damping time of
the fundamental mode of strange stars, letting the parameters of the
bag model vary in the range indicated by Eqs. (\ref{alfas}),
(\ref{ms}) and (\ref{B1}), which covers the parameter space allowed for bare strange stars.
 We consider masses in the range
$[0.7~M_\odot,M_{max}]$, where $M_{max}$ is the maximum mass allowed
by each choice of the model parameters.  We consider bare stars
without a crust, as the presence of a crust does not affect the
fundamental mode frequency in a significant way.

We compare these frequencies with those computed in \cite{astero} for
neutron stars and for hybrid stars.
In \cite{astero} NS were modeled using a set of modern EOS that
describe matter at supranuclear densites; they are
obtained within non relativistic
nuclear many-body theory and relativistic mean field theory, that
model hadronic interactions in different ways, leading to different
composition and dynamics.  The hybrid stars were modeled using
the EOS APR120 and APR200 of Ref. \cite{Akmal},
describing hybrid stars with a rather small admixture of quark matter.

\begin{figure}[ht]
\centering
\includegraphics[angle=270,width=0.6\textwidth]{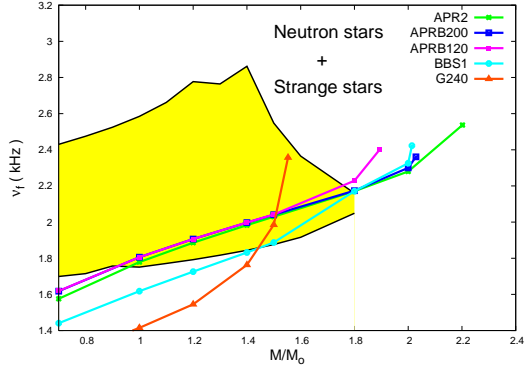}
\caption{The frequency of the fundamental mode is
plotted as a function of the mass of the star, for neutron/hybrid stars
described by the EOS employed in \cite{astero} and indicated with
the same labels (APR2, APRB200, APRB120, BBS1, G240) and for strange
stars.  The shaded region covers the range of parameters of the MIT
bag model considered in this paper, i.e.  $\alpha_s \in [0.4,0.6]$,
$m_s \in [80,155] ~ {\rm MeV}$ and $B \in [57,95]~ {\rm MeV}/{\rm
fm}^3$.  }
\label{fig1}
\end{figure}
\begin{figure}[ht]
\centering
\includegraphics[angle=270,width=0.6\textwidth]{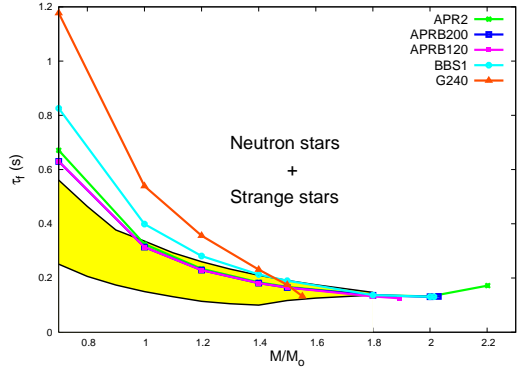}
\caption{ The damping time of the fundamental mode is plotted as a
function of the mass of the star, as in Fig. 1, for neutron/hybrid stars and
for strange stars (shaded region).  }
\label{fig2}
\end{figure}

\begin{figure}[ht]
\centering
\includegraphics[angle=270,width=0.6\textwidth]{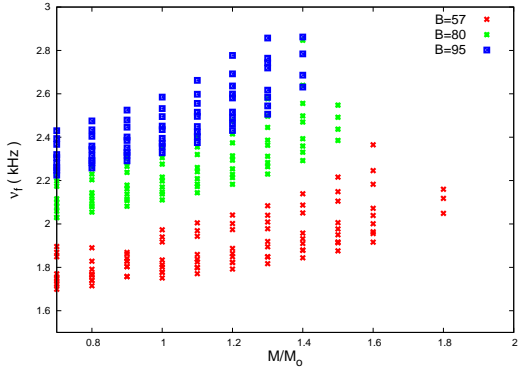}
\caption{The fundamental mode frequency $\nu_f$ is plotted versus the
gravitational mass, $M$, for different values of the bag constant and
$\alpha_s$ and $m_s$ varying in the range indicated by
Eqs. (\ref{alfas}) and (\ref{ms}).  }
\label{fig3}
\end{figure}
\begin{figure}[ht]
\centering
\includegraphics[angle=270,width=0.6\textwidth]{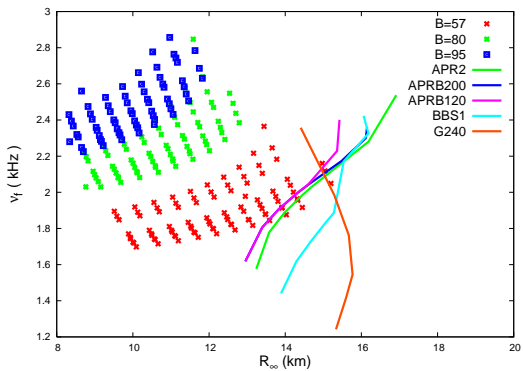}
\caption{$\nu_f$ is plotted versus the radiation radius, $R_\infty$,
for different values of the bag constant and $\alpha_s$ and $m_s$
varying in the range indicated by Eqs. (\ref{alfas}) and (\ref{ms}).
The continuous lines refer to the values of $\nu_f$ for the neutron/hybrid
star models considered in \cite{astero}.  }
\label{fig4}
\end{figure}

The results of this comparison are shown in Figs. 1 and 2, where we
plot $\nu_f$ and the corresponding damping time $\tau_f$,
respectively, versus the mass of the star. In both figures the shaded
region refers to all values of $\nu_f$ and $\tau_f$ allowed for
strange stars, assuming that the paramaters of the bag model vary in
the selected range.  Continuous lines refer to the values of $\nu_f$
and $\tau_f$ computed for neutron and hybrid stars in \cite{astero}.
Figure 1 shows that: 
\begin{itemize}
\item{}
Strange stars cannot emit gravitational waves
with $\nu_f \lesssim 1.7$ kHz for any values of the mass in the range
we consider.
\item{} For masses lower than $1.8~M_\odot$, above which no
stable bare strange star can exist, there
is a small range of frequency where neutron/hybrid stars are
indistinguishable from strange stars.
However, there is a large
frequency region where only strange stars can emit. For instance if
$M=1.2~ M_\odot$, a signal with $\nu_f \gtrsim 1.9$~kHz would belong
to a strange star.
Note that
the fundamental mode frequency and damping time of the hybrid stars 
considered in \cite{Akmal} and \cite{astero} and shown in Figs. 1 and 2 
(EOS APR120 and APR200), are basically indistinguishable from that of the 
neutron star with the same low density EOS, except when
the mass is close to the maximum mass.
This is due  to the small amount of quark matter these hybrid stars contain.
\item{} Even if we do not know the mass of the star
(as it is often the case for isolated pulsars) the knowledge of
$\nu_f$ allows to gain information about the source nature; indeed, if
$\nu_f \gtrsim 2.2$ kHz we can reasonably exclude that the signal is
emitted by a neutron star.
\end{itemize}
Figure 2 contains a complementary information: for strange stars $\tau_f$ is
in general smaller than for neutron/hybrid stars.

Thus, the next question is: assuming that we know the signal has been
emitted by a strange star, can we constrain in some way the parameters
of the MIT bag model ?  In Fig. 3 we show to what extent this is
possible. We plot the values of $\nu_f$ allowed for strange stars
versus the stellar mass, indicating with the same symbol the points
that belong to the same value of the bag constant $B$.  From this
picture we see that for a given mass the mode frequency increases with
$B$, and that, knowing $M$ and $\nu_f$ we would be able to set
constraints on $B$ much more stringent than those provided by the
available experimental data.

Similar information can be derived by the simultaneous knowledge of $\nu_f$
and the radiation radius
\begin{equation}
R_\infty= \frac{R}{\sqrt{1-2M/R}}.
\end{equation}
In Fig. 4 we plot $\nu_f$ as in Fig. 3, but versus $R_\infty$. In
addition we plot as continuous lines the values of $\nu_f$ for the
neutron/hybrid star models considered in \cite{astero}.  The figure shows
that radiation radii smaller than $\simeq 13$~km should be attributed
to strange stars, whereas if $ 13 \lesssim R_\infty \lesssim 15.5 $ Km
the star can either be a strange or a neutron/hybrid star. Higher values of
$R_\infty$ can only belong to neutron/hybrid stars. Figure 4 further shows
that the knowledge of $\nu_f$ constrains the value of $B$.

The problem whether quark stars can be discriminated from neutron
stars using gravitational waves, which we discuss in this letter, has
already been addressed in \cite{yip}, \cite{kojimasakata} and
\cite{Sotani}.

In \cite{yip} $\nu_f$ and $\tau_f$ have been computed for strange star
models obtained within the MIT bag model for two values of $B$, i.e.
$B=56$ and $67$~MeV/fm$^3$, $ m_s=150~ {\rm MeV}$ and assuming 
$\alpha_s$ both vanishing and different from zero.

In \cite{kojimasakata} strange stars have been modeled 
putting $\alpha_s$ and $m_s$  to zero and  choosing 
$B=75$ and $137$~MeV/fm$^3$.

Finally, in \cite{Sotani} the  values of $\nu_f$ and $\tau_f$ have 
been plotted versus the radiation radius for the following
values of the parameters: $\alpha_s=0.6$, $ m_s=0,150, 300 ~ {\rm MeV}$,
and $B=57$ and $209$ MeV/fm$^3$.  These values of the parameters and
the star central density were chosen to fit $R_{\infty}$ within the
range $3.8-8.2$ Km. However, this choice implies very small values of
the stellar mass (ranging from $\sim 0.05$ to $\sim 0.5$~$M_\odot$)
which are hard to explain within current evolutionary scenarios for
neutron stars formation.  As a consequence, they obtain values of
$\nu_f $ larger than $\sim$9~kHz, much higher than those we consider.

The study we propose in this paper differs from the preceeding literature
in what it explores the entire range of allowed parameters in a systematic way.
Our results show that the detection of a signal emitted by a compact
star pulsating in its fundamental mode, combined with a complementary
information on the stellar mass or the radiation radius, would allow
one to discriminate between neutron/hybrid stars and strange stars; in addition,
we show that it would also be possible to
constrain the bag constant to a range much smaller than that provided
by the available data from terrestrial experiments.  

The fundamental mode can be excited in a variety of astrophysical
processes, like in a glitch, in a close interaction with a companion, or after birth
in a gravitational collapse. Recent simulations  of gravitational collapse 
show that a significant fraction of the total energy 
emitted in gravitational waves, of the order of $10^{-9}-10^{-8}~M_\odot c^2$,
is indeed emitted at the  frequency of the $\bf f$-mode
\cite{shibata}, \cite{dimmelm}.
However, this energy is too low to be detectable
by current interferometric antennas like VIRGO or LIGO, unless 
the collapse occurs in our galaxy; but  we know that, unfortunately,
the rate of collapse per galaxy is only of a few per hundred years.
In order to detect signals emitted by more distant sources,
we would need very sensitive, high frequency 
detectors, like EURO or EURO-XYLO,
which have been considered in a preliminary assessment study some years 
ago (http://www.astro.cf.ac.uk/geo/euro/).
As discussed in \cite{cqgvaleria}, this kind of detectors would be 
able to see signals emitted by oscillating stars 
up to the distance of the VIRGO cluster  if 
the energy stored in the mode is of the order of $10^{-7}-10^{-8}~M_\odot c^2$,
which is not too far from present estimates. 
Indeed, current numerical simulations  assume axisymmetric collapse, but if
asymmetries are present the emitted energy may be larger.
And moreover rotation, which is certainly present in stars, has the effect 
of lowering the mode frequencies,  enhancing detection chances.

Thus,  we can conclude that asteroseismology will become a branch 
of gravitational wave research when
high frequency detectors  will be operating, and we
hope that the EURO project will be reconsidered in a not too far future.

It has also to be mentioned that, although somewhat more refined dynamical 
models have been proposed, the MIT bag model
appears to provide a quite reasonable description of quark matter.
The results discussed in Ref. \cite{buballa} show that the EOS obtained
from the Nambu Jona-Lasinio (NJL) model, in which quark masses are
dynamically generated through the appearance of a condensate associated with 
chiral symmetry breaking, is in fact similar to the one obtained from the 
bag model. Using more sophisticated models, such as the NJL model, may turn 
out to be required to describe the possible occurrence of the pairing instability 
induced by the attractive one-gluon exchange interaction between quarks, 
leading to a color superconducting phase \cite{buballa,reddy}.
However, the relative stability of the different superconducting phases discussed 
in the literature is not yet firmly established, and their occurrence is expected to 
affect mostly transport properties and cooling, rather then the stellar structure, and
consequently the $\bf f$-mode frequency, on which we focus in this paper.

\end{document}